# Electron-phonon interaction alone does not explain the photoemission kink in cuprate superconductors

Feliciano Giustino[1,2], Marvin L. Cohen[1,2] & Steven G. Louie[1,2]

[1]*Department of Physics, University of California at Berkeley, California 94720, USA.*

[2]*Materials Sciences Division, Lawrence Berkeley National Laboratory, Berkeley, California 94720, USA.*

**Despite over two decades of intense research efforts, the origin of high-temperature superconductivity in the copper oxides remains elusive. Angle-resolved photoemission experiments[1,2] (ARPES) revealed a kink in the dispersion relations (energy vs. wavevector) of electronic states in the cuprates at binding energies of 50-80 meV, and raised the hope that this anomaly could be key to understanding high-temperature superconductivity. The kink is often interpreted in terms of interactions between the electrons and a bosonic field. While there is no consensus on the nature of the bosons or even whether a boson model is appropriate, phonons[1] and spin fluctuations[2] have alternatively been proposed as possible candidates. Here we report state-of-the-art first-principles calculations of the role of phonons and the electron-phonon interaction in the photoemission spectra of $La_{2-x}Sr_xCuO_4$ (LSCO). Our study demonstrates that the phonon-induced renormalization of the electron energies and the Fermi velocity is almost one order of magnitude smaller than the effect observed in photoemission experiments. Hence, the present finding rules out electron-phonon interaction in bulk LSCO as the possible origin of the measured kink. This result should bear on several proposed theories of high temperature**

**superconductivity in addition to theories concerning the origin of kinks in cuprate photoemission data.**

We studied LSCO since it crystallizes in a simple lattice characterized by a single $CuO_2$ layer, and experimentally the hole concentration can be controlled over the entire phase diagram through the Sr content[3]. We considered optimally doped ($x = 0.15$) and heavily overdoped ($x = 0.30$) LSCO for which a mean-field description of the electronic structure such as that given by density functional theory is appropriate to interpret measured Fermi surfaces[4]. We did not attempt to address the electron-phonon coupling in the underdoped compound since in that case the validity of the Fermi liquid paradigm has been questioned[5]. In order to calculate the electronic structure of LSCO we employed a generalized gradient approximation to density functional theory[6,7] which correctly predicts the antiferromagnetic ground state of the parent oxide with no adjustable parameters (see Supplementary Information). The calculated Fermi surfaces of LSCO are generally in good agreement with ARPES experiments[8] (cf. Supplementary Figure 1), and in particular they reproduce the evolution of the Fermi surface topology from a (0, 0)-centered electron-pocket in overdoped LSCO to a ($\pi$, $\pi$)-centered hole-pocket in the underdoped compound[9] (wavevectors are given in units of $1/a$, $a = 3.77$ Å being the lattice parameter).

We studied the lattice dynamics of LSCO by calculating the phonon eigenmodes and eigenenergies across the entire Brillouin zone through density functional perturbation theory[10]. In Fig. 1 we compare the calculated phonon dispersions along the antinodal direction (0,0)-($\pi$,0) and the total vibrational density of states to inelastic neutron scattering data[11,12]. The dispersion of the in-plane half-breathing Cu-O stretching mode is found to be in very good agreement with experiment. Inspection of the vibrational density of states shows that a one-to-one correspondence can be established between calculated and measured peaks, indicating an overall agreement between theory and experiment[11,12,13,14].

An electron traveling in a solid distorts the lattice due to the Coulomb interaction with the ions. The lattice distortion in turn has a feedback on the electron dynamics, resulting in an increase of the electron mass and a shortening of the electron lifetime in a particular quasiparticle state. Within quantum field theory, this effect can conveniently be described in terms of a complex self-energy $\Sigma$ that the electron acquires as a consequence of the electron-phonon interaction. The real part of the self-energy describes the change in the electron energy due to this interaction, while the imaginary part provides information on the electron lifetime $\tau$ through $\tau = \hbar / (2\,\mathrm{Im}\,\Sigma)$. We calculated the electron self-energy within the Migdal approximation to the Feynman-Dyson perturbation theory[15,16,17]. Due to the $d$-wave symmetry of the pseudogap[18] and of the superconducting gap[19], this is a reasonable approximation for the $(0, 0)$-$(\pi, \pi)$ nodal direction. By focusing on the nodal direction we also avoid the van Hove singularity at $(\pi, 0)$ and the comparison with experiment can be performed with little ambiguity. The diagonal part of the electron self-energy operator $\Sigma(\mathbf{r}, \mathbf{r}', \omega)$ with respect to the unperturbed electronic states arising from the electron-phonon interaction reads[15,16,17]:

$$
\begin{aligned}
\Sigma_{n\mathbf{k}}(\omega) = \left\langle n\mathbf{k} \left| \Sigma(\mathbf{r}, \mathbf{r}', \omega) \right| n\mathbf{k} \right\rangle = \sum_{m\nu} \int \frac{d\mathbf{k}'}{\Omega_{\mathrm{BZ}}} \left| g_{mn,\nu}(\mathbf{k}, \mathbf{k}') \right|^2 \times \\
\times \left[ \frac{1 - f_{m\mathbf{k}'} + n_{\mathbf{q}\nu}}{\omega - \varepsilon_{m\mathbf{k}'} - \omega_{\mathbf{q}\nu} - i\delta} + \frac{f_{m\mathbf{k}'} + n_{\mathbf{q}\nu}}{\omega - \varepsilon_{m\mathbf{k}'} + \omega_{\mathbf{q}\nu} - i\delta} \right]
\end{aligned}
\tag{1}
$$

where $|n\mathbf{k}\rangle$ indicates a Bloch eigenstate with wavevector $\mathbf{k}$, band index $n$ and energy $\varepsilon_{n\mathbf{k}}$, $g_{mn,\nu}(\mathbf{k}, \mathbf{k}') = \left\langle m\mathbf{k}' \left| \Delta_{\mathbf{q}\nu} \mathrm{V}(\mathbf{r}) \right| n\mathbf{k} \right\rangle$ is the electron-phonon matrix element for the

scattering $|n\mathbf{k}\rangle \rightarrow |m\mathbf{k}'\rangle$ through a phonon of wavevector $\mathbf{q} = \mathbf{k} - \mathbf{k}'$, branch index $\nu$ and energy $\omega_{\mathbf{q}\nu}$. The operator $\Delta_{\mathbf{q}\nu}V(\mathbf{r})$ represents the variation of the self-consistent potential with respect to a collective lattice displacement associated with this phonon, and the fermion and boson occupations $f_{m\mathbf{k}'}$ and $n_{\mathbf{q}\nu}$ account for the temperature dependence of the self-energy. The accurate evaluation of the electron self-energy requires the calculation of the electron-phonon matrix elements $g_{mn,\nu}(\mathbf{k},\mathbf{k}')$ on an extremely fine mesh consisting of more than a hundred thousand phonon wavevectors. This represents a formidable computational task and was beyond the reach of existing approaches. In order to overcome this difficulty we developed a new technique based on the Wannier representation[20] which exploits the extreme short-range of the electron-phonon interaction in real space[21,22]. Figure 2 shows the electron self-energy calculated for optimally doped and overdoped LSCO at 20 K. At optimal doping the real part of the self-energy displays a peak at a binding energy of 70 meV, a shoulder at 40 meV, and a broad background extending up to much higher binding energies. A mode-resolved analysis shows that the main peak at 70 meV can be assigned to the Cu-O in-plane half-breathing motion around ($\pi$, 0) and full breathing motion around ($\pi$, $\pi$). The shoulder at 40 meV arises from phonons involving components of both out-of-plane buckling of the planar O atoms and in-plane O-O stretching vibrations. The broad high-energy background can be shown to arise from a density-of-states effect (cf. caption of Fig. 2). As a consequence of the larger hole concentration, in the overdoped regime the high-energy background moves closer to the Fermi level, whereby the main peak appears slightly blue-shifted and the shoulder looses intensity.

In order to clarify the origin of the 70 meV peak and the 40 meV shoulder we plot in Fig. 3 the squared electron-phonon matrix elements $|g(\mathbf{k},\mathbf{k}')|^2$, which represents the transition probability from a initial Cu $d_{x^2-y^2}$-O $p_\sigma$ nodal state with wavevector $\mathbf{k}$ on the

Fermi surface to a final state $\mathbf{k'}$, associated with a given phonon $\mathbf{q} = \mathbf{k'} - \mathbf{k}$. Our calculations show that the matrix elements corresponding to the half-breathing and the full breathing modes are by far the largest among all vibrational modes (with the exception of the apical O stretching mode, cf. caption of Fig. 3), thereby explaining the main peak at 70 meV in the self-energy. On the other hand, the 40 meV buckling/stretching modes show only moderately large matrix elements as compared to the 70 meV branches. However, these modes connect the nodal region with a large portion of the Fermi surface, and therefore their contribution to the self-energy is enhanced by a significant phase-space effect.

We now proceed to a quantitative comparison between our calculations and the photoemission data of ref 1. The measured peaks provide the quasiparticle energy $E_{\mathbf{k}}$ for a given parallel momentum $\mathbf{k}$. The real part of the self-energy can be extracted from $\mathrm{Re}\,\Sigma_{\mathbf{k}}(E_{\mathbf{k}}) = E_{\mathbf{k}} - \varepsilon_{\mathbf{k}}$, with $\varepsilon_{\mathbf{k}}$ the non-interacting energy[15]. Along the nodal direction the non-interacting dispersions are linear and we can assume $\varepsilon_{\mathbf{k}} = \mathbf{v} \cdot \mathbf{k}$ as commonly done in the literature, where the bare velocity $\mathbf{v}$ is to be determined. We determined the bare velocity by fitting the raw data of Figs. 1(a) and (d) of ref 1 to a simple Holstein self-energy model (see Supplementary Information). With the experimentally determined $\mathbf{v}$, the real part of the self-energy is extracted and shown in Fig. 4. A visual comparison between the theoretical self-energy and the experimental data indicates that the effect predicted by theory is considerably weaker. We note that this result does not depend on the choice of the bare band dispersions. Indeed, when comparing the raw data of ref 1 with our calculated electron dispersions fully renormalized by the electron-phonon interaction [ $E_{\mathbf{k}} = \varepsilon_{\mathbf{k}} + \mathrm{Re}\,\Sigma_{\mathbf{k}}(E_{\mathbf{k}})$, panels (b) and (c) of Fig. 4], the discrepancy between experiment and theory is unmistakably visible. In order to be quantitative and avoid ambiguity, we extracted the electron-phonon coupling strength $\lambda$ by taking the low-energy slopes of both the theoretical and the experimental self-energy data sets. The rationale for this procedure is that, in the small binding energy limit, the self-energy arising from the electron-phonon interaction reads $\mathrm{Re}\,\Sigma(\omega) \approx -\lambda\,\omega$. We took the slope at the Fermi level as an upper bound to $\lambda$, and the slope between the Fermi level and a binding energy of 50 meV as a lower bound. This procedure yielded the average electron-phonon couplings from experiment $\lambda_{\mathrm{expt}}$ =1.00-1.32 for the optimally doped sample at 20 K, and $\lambda_{\mathrm{expt}}$ = 0.75-0.99 for the overdoped sample at 20 K, consistent with the estimates provided in ref 1. From the calculated self-energy, our analysis however yielded $\lambda_{\mathrm{th}}$ = 0.14-0.22 at optimal doping and $\lambda_{\mathrm{th}}$ =0.14-0.20 in the overdoped regime. Hence, the calculated renormalization of the nodal Fermi velocity is $\lambda_{\mathrm{expt}}/\lambda_{\mathrm{th}}$ = 5-7 times smaller than in experiment.

The discrepancy between theory and experiment is so large that it is unlikely to arise from any uncertainties in the calculations within our theoretical framework (see Supplementary Information) or from experiment. It is important to note that this disagreement is also found in the overdoped regime where the effects of electron correlations are reduced. These observations lead us to conclude that the electron-phonon interaction in bulk LSCO cannot account for the experimentally observed kink.

In retrospect, our results are consistent with previous first-principles calculations[23,24] on the electron-phonon interaction in the cuprates although these studies did not address the ARPES anomaly discussed here. Brillouin zone-averaged coupling strengths $\lambda_{\mathrm{ave}} \sim 0.4$ and $\lambda_{\mathrm{ave}}$ = 0.27 have been obtained using linear-response techniques for $Ca_{0.27}Sr_{0.63}CuO_2$ and $YBa_2Cu_3O_7$, respectively[23,24]. When we evaluate the same parameter using the standard definition, corresponding to the first reciprocal energy moment of the Eliashberg function[15] $\alpha^2F(\omega)$, we obtain $\lambda_{\mathrm{ave}}$ = 0.4 for optimally doped

LSCO, in agreement with those studies. Furthermore, a recent theoretical work[25] on the electron self-energy in $YBa_2Cu_3O_7$ reports an electron-phonon contribution whose magnitude is consistent with our results.

The analysis of alternative scenarios involving spin excitations[26], such as the magnetic resonance mode and the electron-magnon interaction, as well as other possible effects[27], is beyond the scope of the present work. However, the contribution of non-phononic mechanisms to the kink remains an important open question. We wish to stress that the present study focuses on the photoemission kink and does not rule out theories of high-temperature superconductivity based on novel phonon mechanisms and topological approaches, such as the dopant percolative model[28].

**Supplementary Information** accompanies the paper on **www.nature.com/nature**.

**Acknowledgements**

The authors thank Y.-W. Son and C.-H. Park for fruitful discussions. This work was supported by the National Science Foundation and by the Director, Office of Science, Office of Basic Energy Sciences, Materials Sciences and Engineering Division, U. S. Department of Energy. Computational resources were provided by NPACI and NERSC. Part of the calculations were performed using the WANNIER [A. Mostofi *et al.*, www.wannier.org] and the Quantum-ESPRESSO [S. Baroni *et al.*, www.pwscf.org] packages. The Fermi surfaces were rendered using XCRYSDEN [A. Kokalj, www.xcrysden.org].

Correspondence and requests for materials should be addressed to sglouie@berkeley.edu.

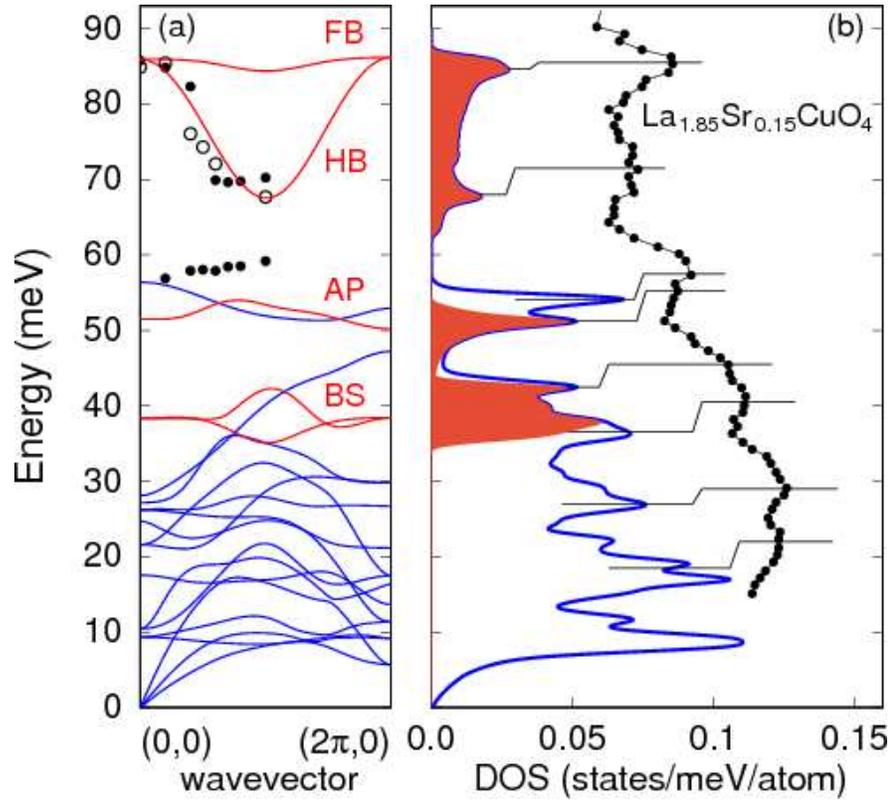

**Figure 1. Phonons of La$_{2-x}$Sr$_x$CuO$_4$ at optimal doping**. (a) Calculated dispersions along the antinodal direction (0, 0)-(2π, 0) (solid lines), compared to the inelastic neutron scattering data of ref 11 (circles: 300 K data, disks: 10 K data). The labels indicate the phonon branches associated with the full-breathing (FB) in-plane Cu-O stretching mode at (π, π), to the half-breathing (HB) in-plane Cu-O stretching mode at (π,0), the *c*-axis apical (AP) O stretching mode at (0,0), and the buckling/stretching (BS) modes at (π,0) and (π, π). (b) Calculated phonon density of states (solid line), compared to the inelastic neutron scattering data of ref 12 (disks). The broken lines indicate the correspondence between measured and calculated peaks. The theoretical phonon energies agree with experiment to within 5 meV. The shaded regions correspond to the phonon branches highlighted in (a). The calculated half-breathing (π,0) and apical (0,0)

frequencies are 68 meV and 53 meV, compared to the corresponding experimental values of 68-70 meV and 57 meV[11], respectively. In order to extend the present comparison to the undoped parent oxide, we also performed the calculations for the $(\pi, 0)$ half-breathing mode in antiferromagnetic $La_2CuO_4$. The theoretical energy of 76 meV agrees with neutron scattering measurements yielding 74 meV[13], and indicates that our theoretical framework correctly describes the softening of the half-breathing phonon through the insulator-to-metal transition[12,13,14].

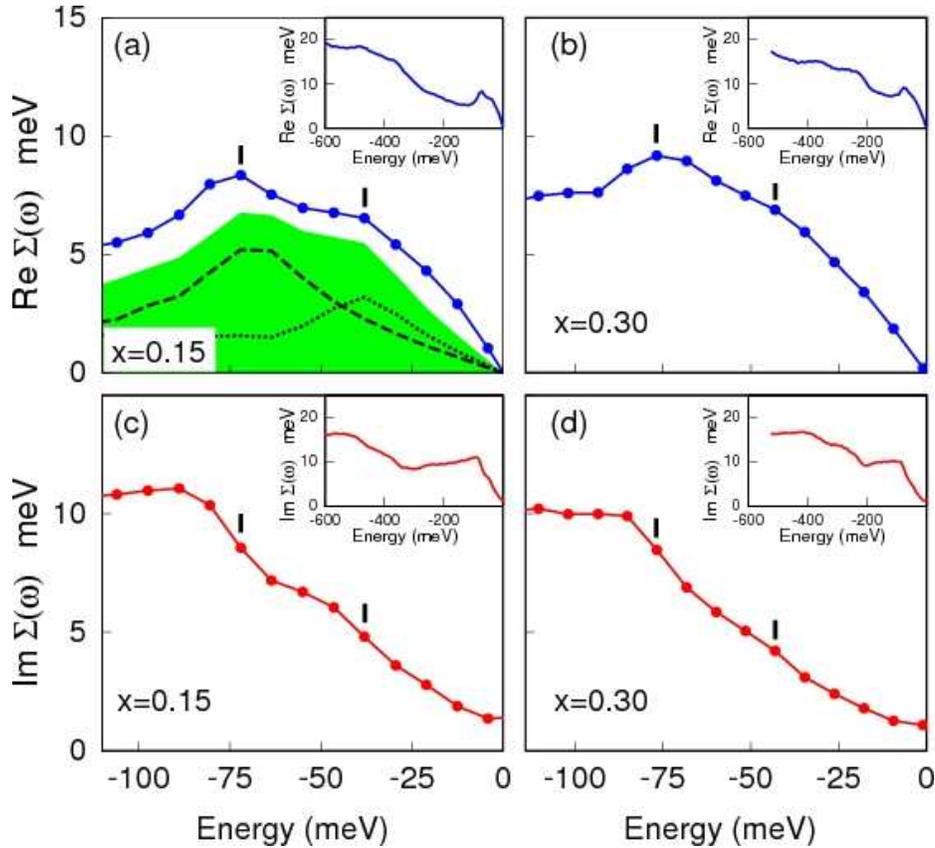

**Figure 2. Calculated electron self-energy in LSCO due to the electron-phonon interaction**. (a), (b) Real part of the self-energy for optimally doped ($x$ = 0.15) and overdoped ($x$ = 0.30) LSCO, respectively, at the temperature T = 20 K. (c), (d) Imaginary part of the self-energy for optimally and overdoped LSCO. Insets: real and imaginary part of the self-energy on an extended energy scale. We have computed the self-energy for electrons with the parallel momentum **k** along the cut $(0,0,k_\perp)$-$(\pi,\pi,k_\perp)$ [see Supplementary Figure 1(b)] for three different values of the normal component $k_\perp$ (0, $\pi a/c$, and $2\pi a/c$). The variation of the self-energy along the $c$-axis is negligible along the cuts studied, therefore we only show the results for the cut through the zone-center. The black markers in (a) and (b) indicate the main peak at ~70 meV and the shoulder at ~40 meV. The black dashed and dotted lines represent the contributions to the self-energy from the half-breathing and full-breathing Cu-O stretching modes, and from the breathing/stretching modes around 40 meV, respectively. The sum of these two contributions (green area) accounts for more than 80% of the real part of the self-energy. The imaginary part of the self-energy in the insets of (c) and (d) exhibits a step-like behaviour with a leading edge around 40-70 meV. This is a consequence of Pauli's exclusion principles: holes with energy below the threshold for phonon emission ~40-70 meV cannot make transitions to above the Fermi level emitting a phonon and thus exhibit longer lifetimes. Since the electronic density of states increases with increasing binding energy, a hole with large binding energy has an increased probability of decaying through phonon emission/absorption due to the larger number of final states available. The increased decay probability is reflected in the imaginary part of the self-energy and gives rise to the broad high-energy background in the real part through a Kramers-Krönig relation (see Supplementary Discussion and Supplementary Figure 2).

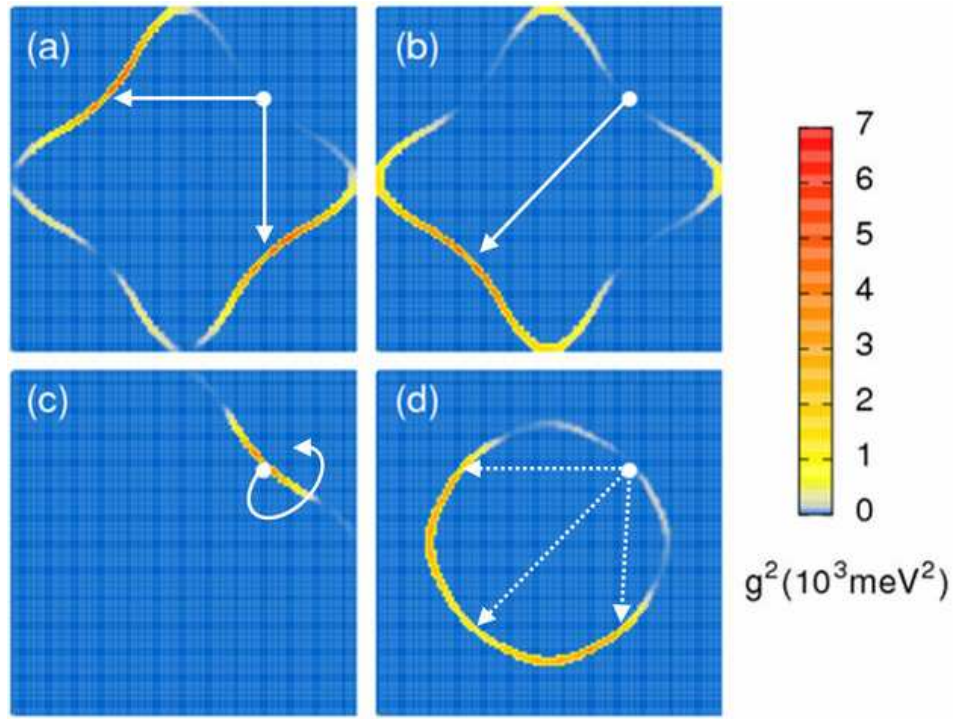

**Figure 3. Electron-phonon matrix elements.** Intensity plots of the squared electron-phonon matrix elements $|g(\mathbf{k}, \mathbf{k'})|^2$ giving the transition probability for the scattering from a state with $\mathbf{k}$ on the Fermi surface (indicated by the white dot) to final states $\mathbf{k'}$ on the Fermi surface, for optimally doped LSCO. (a) The matrix elements associated with the half-breathing phonons connect adjacent Fermi surface arcs with a maximum strength $g^2 = 7.0 \cdot 10^3$ meV$^2$. (b) The full-breathing modes connect Fermi arcs on the opposite sides of the Fermi surface with a maximum strength $g^2 = 6.7 \cdot 10^3$ meV$^2$. (c) The apical O stretching mode also shows large matrix elements $g^2 = 6.5 \cdot 10^3$ meV$^2$. However, this phonon only gives rise to momentum-conserving transitions $\mathbf{k} = \mathbf{k'}$, hence the number of allowed final states is negligible and its coupling to nodal holes is frustrated. (d) The squared matrix element for the mixed buckling/stretching modes at 40 meV. These modes involve significant momentum transfer along the $c$-axis ($k'_\perp - k_\perp =$

$2\pi a/c$) and connect the nodal region to a large portion of the Fermi surface. The largest matrix elements are found at the intersection of the Fermi surface with the $(k_x, k_y, \pm 2\pi a/c)$ planes. The largest matrix element in this case is $g^2 = 4.3 \cdot 10^3$ meV$^2$. The squared matrix elements in overdoped LSCO are similar to those obtained at optimal doping (within 20%).

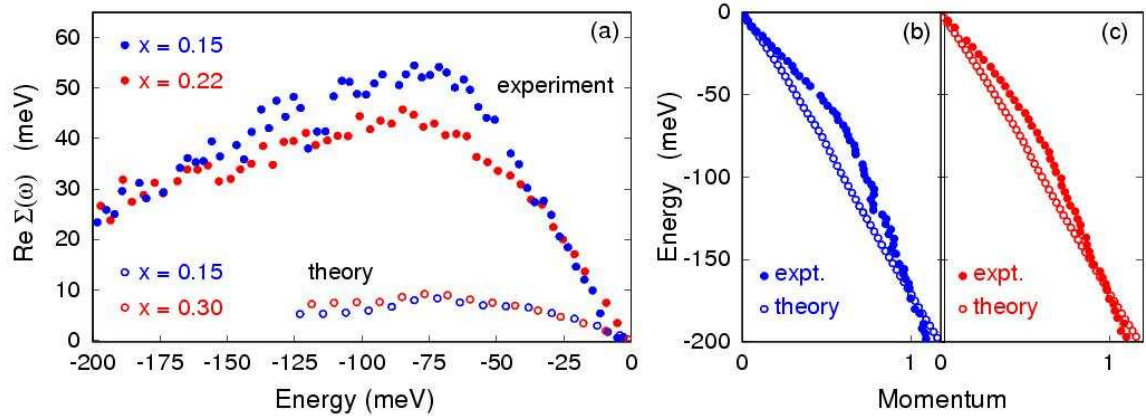

**Figure 4. Comparison between theory and experiment.** (a) Comparison between the real part of the electron self-energy as obtained by ARPES experiments[1] and the first-principles theoretical self-energy arising from the electron-phonon interaction. From the raw data of Figs. 1(a) and (d) of ref 1 we extracted the self-energy assuming a linear dispersion and using a bare velocity $|\mathbf{v}|$ = 3.4 eV ·Å obtained by a fit to the data with an Holstein electron-phonon self-energy model (see Supplementary Discussion and Supplementary Figure 3, the precise value of the bare velocity is not critical to our conclusions). We show the experimental data at 20 K for optimally doped LSCO ( $x$ = 0.15) and for overdoped LSCO ( $x$ = 0.22), as well as the calculated self-energy at optimal doping ( $x$ = 0.15) and in the overdoped case ( $x$ = 0.30) at the same temperature. (b), (c) Comparison between the quasiparticle dispersion relations obtained from the peaks of the photoemission spectra and our calculations including the renormalization due to the electron-phonon interaction [(b) optimally doped, (c) overdoped]. We followed ref 1 in normalizing the photoelectron wavevector to 1 at the binding energy of 170 meV. The magnitude of the calculated self-energy is so small that the fine structures corresponding to the 40 meV and 70 meV excitations shown in Fig. 2 are barely seen. The Migdal

approximation to the Feynman-Dyson perturbation theory [Eq. (1)] using the density functional theory band structure is reasonable for states along the nodal direction in the normal, superconducting and pseudogap phases because the phonons which are responsible for the 70 meV peak in Fig. 2(a) mainly connect the nodal regions among themselves [Fig. 3(a) and (b)]; therefore, the important transitions contributing to the self energy are those among the Fermi surface arcs, where the gap function is very small.

# Supplementary Information

# Electron-phonon interaction alone does not explain the photoemission kink in cuprate superconductors

Feliciano Giustino, Marvin L. Cohen, and Steven G. Louie

## Supplementary Discussion

**Computational details of electronic structure and self energy**

We described the electronic structure of LSCO within the generalized gradient approximation (GGA) of Perdew and Wang to density functional theory[1]. This functional has the advantage of correctly predicting an antiferromagnetic insulating ground state for the parent oxide $La_2CuO_4$. Indeed, test calculations of $La_2CuO_4$ in the orthorhombic structure indicate that this functional leads to a spin polarized solution with an antiferromagnetic moment of 0.58 $\mu_B$, in good agreement with the experimental value[2] of 0.5 $\mu_B$. The antiferromagnetism lifts the degeneracy between the two Cu sites and results in a small indirect band gap of 0.23 eV. This finding is consistent with previous studies indicating that, within the local spin-density approximation, $La_2CuO_4$ is at the edge of an antiferromagnetic phase transition[3]. The underestimation of the experimental band gap of $La_2CuO_4$ corresponds to a well-known deficiency of the density functional theory[4,5].

The valence electronic wavefunctions were expanded in a planewaves basis[6] with a kinetic energy cutoff of 80 Ry. The core-valence interaction was taken into account by means of norm-conserving pseudopotentials of the Troullier-Martins type[7], including the $s$ and $p$ semicore states for La and the nonlinear core-correction[8] for Cu. The lattice-dynamical calculations for the antiferromagnetic parent oxide were performed using a grid with 8×8×2 points in the orthorhombic Brillouin zone (84 irreducible points). The convergence of the antiferromagnetic moment was checked using up to 12×12×12 points (corresponding to 504 irreducible points). In all our calculations we employed the fully relaxed tetragonal cell (space group I4/mmm) with parameters $a$=3.780 Å, $c/a$=3.441, $z_{La}/c$=0.362, $z_{O2}/c$=0.184. These parameters compare well with neutron power diffraction analysis[9] yielding $a$=3.765 Å, $c/a$=3.505, $z_{La}/c$=0.361, $z_{O2}/c$=0.183 for $x$=0.15 and $a$=3.756 Å, $c/a$=3.522, $z_{La}/c$=0.360, $z_{O2}/c$=0.181 for $x$=0.30. Although at ambient pressure and 10 K the tetragonal-to-orthorhombic phase transition takes place at $x = 0.20$, we described optimally doped LSCO using a tetragonal cell. This is a sensible simplification since studies of LSCO under pressure showed that at 20 Kbar the structure remains tetragonal for all Sr concentrations, and at the same time the shape and magnitude of the superconducting dome is essentially identical to that measured at ambient pressure[10]. We simulated the Sr doping by removing electrons and adding a neutralizing background charge[11].

The Wannier-Fourier interpolation[12,13] of the electronic eigenvalues and eigenstates, the phonon energies and eigenmodes, and the electron-phonon matrix elements was performed starting from a 4×4×4 Brillouin zone coarse mesh and checking the convergence with test calculation using an 8×8×8 coarse mesh. In order to achieve numerical convergence, we had to include up to 128,000 inequivalent **q**-points in the fine mesh used for the integration of the electron self-energy (Fig. 2). For the analysis of the anisotropy and strength of the electron-phonon matrix elements (Fig. 3) we adopted an even finer grid of 450,000 **q**-points. The calculations of the electron self-energy were performed by including 17 valence bands. In terms of Wannier functions, these bands are associated with the 5 Cu-$3d$ states and the 12 O-$2p$ states. In order to check the convergence of the electron self-energy with respect to the number of electronic bands, we performed part of the calculations using an extended set of 29 bands including the 10 La *5d* states. Overall, we considered an energy range of ±10 eV around the Fermi level for the virtual transitions in the real part of the electron self-energy. The smearing δ in Eq. (1) was chosen to be 1 meV, slightly larger than the accuracy of our calculated phonon energies.

We checked the convergence of the kinetic energy cutoff by repeating the calculations of the phonon dispersions with a cutoff of 120 Ry. The corresponding variation of the phonon energies was found to be negligible (0.5 meV at most).

**High-energy background in the electron self-energy**

In order to show that the broad high-energy background in the calculated self-energy in Fig. 2 arises from a density-of-states effect, we rewrite Eq. (1) by assuming a Holstein model, corresponding to an Einstein phonon spectrum peaked at the frequency $\Omega$ and a constant electron-phonon matrix element $g$. We also consider the zero-temperature limit for simplicity. For the imaginary part of the self-energy we find in this model:

$$\mathrm{Im}\,\Sigma(\omega) = \begin{cases} \pi\,g^2\,N(\omega - \Omega\,\mathrm{sgn}\,\omega) & |\omega| > \Omega \\ 0 & |\omega| < \Omega \end{cases}, \tag{S1}$$

$N$ being the electronic density of states. The corresponding real part of the self-energy is found using the Kramers-Krönig relations. The self-energy calculated within this simplified model with $\Omega = 70$ meV and $g = 75$ meV is shown in Supplementary Figure 2. The close match between the first-principles calculation and the model allows us to assign the high-energy background to a density of states effect.

**Extraction of the self-energy from experiment**

We determined the noninteracting quasiparticle velocity $|\mathbf{v}|$ by fitting the raw data of Figs. 1(a) and (d) of ref 14 to the simplest self-energy model based on a single Einstein oscillator and a constant density of states[15] (usually referred to as "Holstein model"):

$$E_{\mathbf{k}} = \mathrm{Re}\,\Sigma(E_{\mathbf{k}}) + \mathbf{v} \cdot \mathbf{k} \,, \tag{S2}$$

$$\mathrm{Re}\,\Sigma(\omega) = -\frac{1}{2}\Omega\lambda\log\left|\frac{\omega + \Omega - i\Gamma}{\omega - \Omega - i\Gamma}\right|, \tag{S3}$$

Besides the bare velocity, the free parameters in the fitting are the characteristic phonon energy $\Omega$, the coupling strength $\lambda$ and the broadening $\Gamma$. Such simplified model can be derived from the Holstein model of Eq. (S1) by assuming that the electronic density of states is constant and equal to the value at the Fermi level. Supplementary Figure 3 shows the self-energy determined in this way from the data of ref 14. The good quality of the fits indicates that the procedure adopted here is meaningful. This analysis gives for the model parameters the values $\Omega = 63$ meV (73 meV), $\lambda = 1.49$ (1.02), and $\Gamma = 58$ meV (37 meV) for the sample with $x = 0.15$ ($x = 0.22$) at 20 K. The bare Fermi velocity is found to be 3.38 eV·Å (3.40 eV·Å) for the optimally doped (overdoped) sample, in reasonable agreement with our calculated GGA velocity 3.8 eV·Å. The fact that the velocity extracted by using the Holstein model exceeds our GGA velocity is consistent with the recent discovery of a high-energy kink at 300-400 meV binding energy which introduces an additional smooth contribution to the self-energy at low binding energy[16,17]. We note that the precise value of the noninteracting Fermi velocity is not crucial to any of our conclusions.

**Discussion of the approximations adopted**

We here discuss the theoretical approximations adopted in the calculation of the self-energy.

*(i) The calculations were performed within the harmonic approximation for the phonons.* It has been shown that only two phonon modes are anharmonic, the tetrahedral tilt mode and the sliding mode of the La atoms in the blocking layer[18]. These modes are soft[19] around $(\pi, \pi)$ and their contribution to the self-energy is not included in our calculations. While it cannot be excluded that such modes may increase the electron self-energy, if their contribution was significant the kink should be observed at a binding energy much smaller than 70 meV.

*(ii) A Fermi-level shift and a neutralizing background were employed to account for the Sr doping.* A previous calculation of $La_{2-x}Sr_xCuO_4$ taking into account the random alloying with Sr within the coherent potential approximation[20] showed that the electronic structure close to the Fermi level follows a rigid-band behavior. In addition, a calculation of $LaBaCuO_4$ within the linearized augmented planewave method found that a rigid-band behavior describes well the electronic structure near the Fermi level[21]. Besides, our computed Fermi surfaces are in good agreement with the most recent measurements. All of these results justify our modeling of the Sr doping. The main drawback of the present approximation is that localization effects associated with the dopant are not taken into account. However, while hole-doping in the blocking layer (as opposed to the copper oxide planes) has recently been suggested theoretically[22], experimental evidence does not appear to support this possibility[23].

*(iii) We adopted a mean-field description of the electronic structure.* While such a level of theory is not adequate in underdoped LSCO, it has been demonstrated that a good agreement is found between photoemission maps and local density calculations for optimally and overdoped LSCO when matrix element effects and three-dimensionality are taken into account[24].

In particular, in the overdoped regime the low energy quasiparticle dispersions obtained within the local density approximation are in good agreement with recent photoemission experiments[25], and in overdoped LSCO our calculated self-energy is still significantly smaller than in experiment.

**Supplementary References**

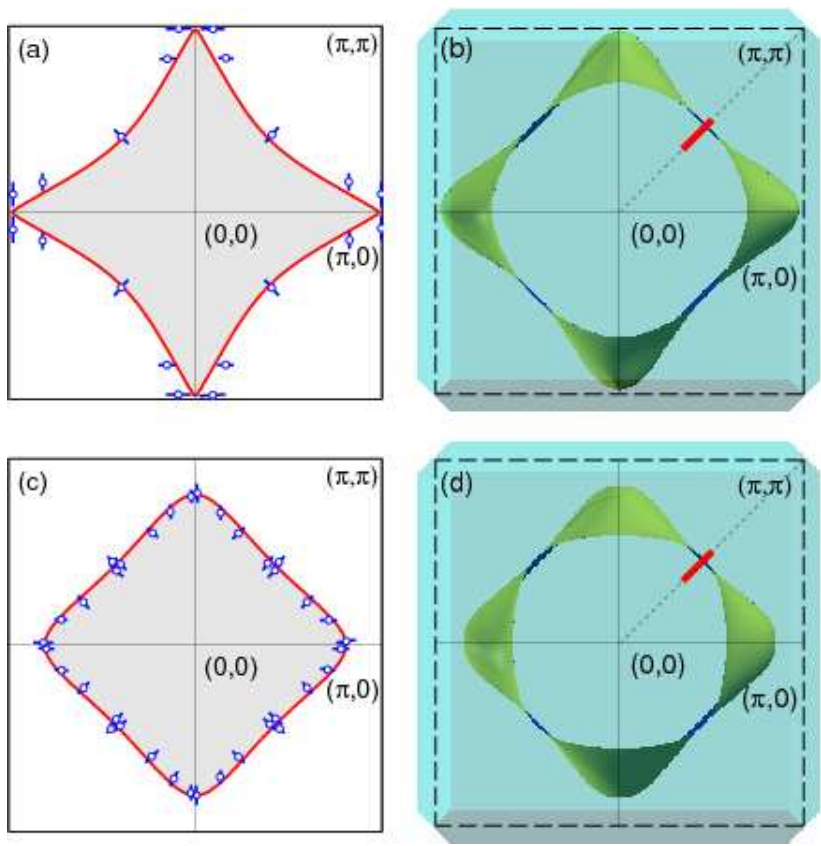

**Supplementary Figure 1**. **Fermi surfaces of La$_{2-x}$Sr$_x$CuO$_4$**. Photoemission data from ref 27 for *x* = 0.15 (a), and *x* = 0.30 (c), and theoretical Fermi surfaces for *x* = 0.15 (b), and *x* = 0.30 (d). The experimental Fermi surface crossings (blue segments) correspond to the wavevectors where the leading edge of the photoemission intensity has a local maximum (the width of the error bars indicates the momenta where most of the spectral weight is below the Fermi level and is almost negligible above the Fermi level, respectively. From ref 26). The calculated Fermi surfaces in (b) and (d) are shown in a top view of the three-dimensional Brillouin zone of LSCO (green). The spread in the plots reflects the c-axis dispersion of the Fermi surfaces. In the nodal regions the the innermost

surface (blue) corresponds to $k_z = 0$ and the outermost surface (green) to $k_z = 2\pi/c$. The calculated nodal Fermi surface crossings 0.46 ($\pi,\pi$) and 0.44 ($\pi,\pi$) for optimally doped and overdoped LSCO compare well with the measured Fermi momenta 0.41±0.04 ($\pi,\pi$) and 0.42±0.04 ($\pi,\pi$), respectively. The comparison is less favorable for the antinodal Fermi surface crossing at optimal doping. Both the experimental and the theoretical determination of the antinodal crossing are complicated by the saddle-point van Hove singularity at ($\pi$, 0) and the non-negligible $c$-axis dispersion. In (b) and (d) we show as red segments the nodal cuts along which we calculated the electron self-energy.

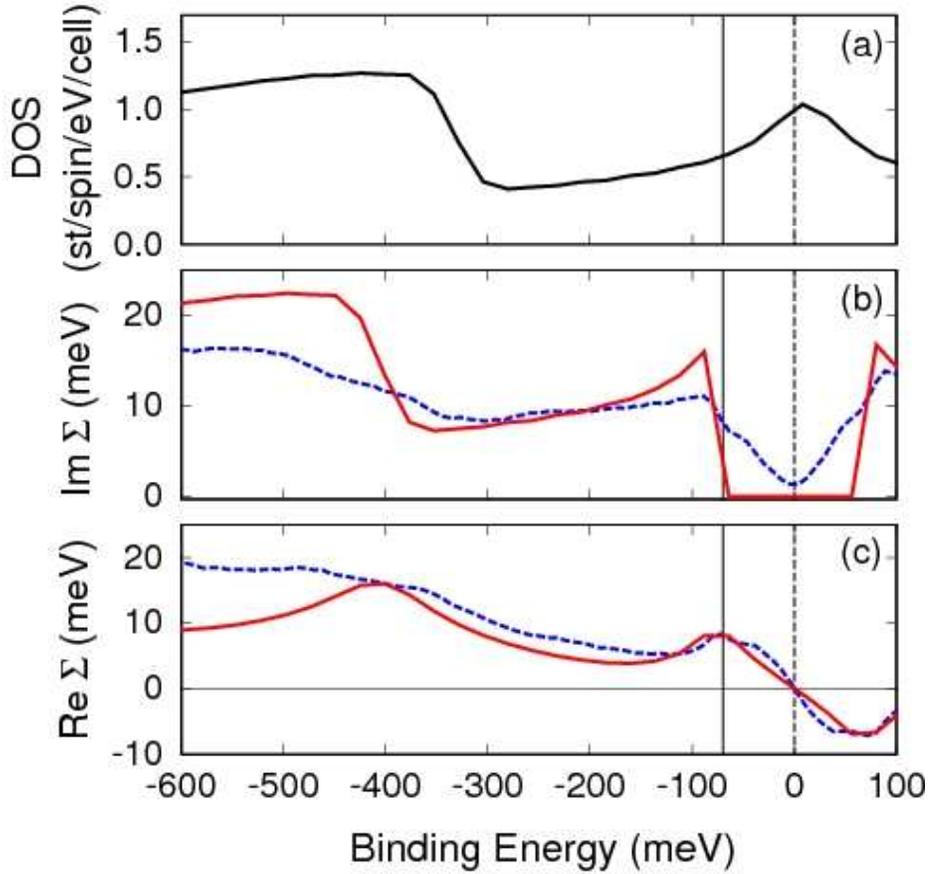

**Supplementary Figure 2. Electron self-energy arising from the electron-phonon interaction in optimally doped LSCO**. (a) Density of electronic states. The peak at the Fermi level corresponds to the saddle-point van Hove singularity of the Cu $d_{x^2-y^2}$ -O $p_\sigma$ band at the antinode ($\pi$, 0). The step at the binding energy ~330 meV corresponds to the local maxima of the Cu $d_{z^2}$ -O $p_z$ band at ($\pi$,$\pi$). The binding energy of this feature may be underestimated by as much as 0.5 eV in our calculations[27]. (b) Imaginary part of the electron self-energy: *ab-initio* calculation (dashed blue line) and Holstein model [Eq. (S1), solid red line] with $\Omega$ = 70 meV and *g* = 75 meV. In the Holstein model, the imaginary part of the self-energy is obtained from the density of states by "opening a gap" 2 $\Omega$ at the Fermi level. (c) Real part of the electron self-energy: *ab-initio* calculation (dashed

blue line) and Holstein model [solid red line, obtained from the imaginary part in (b) through the Kramers-Krönig relations]. The *ab-initio* self-energy appears broader than the Holstein model due to the inclusion of all phonon modes and the variation of the electron-phonon matrix element with the energy of the final electronic state.

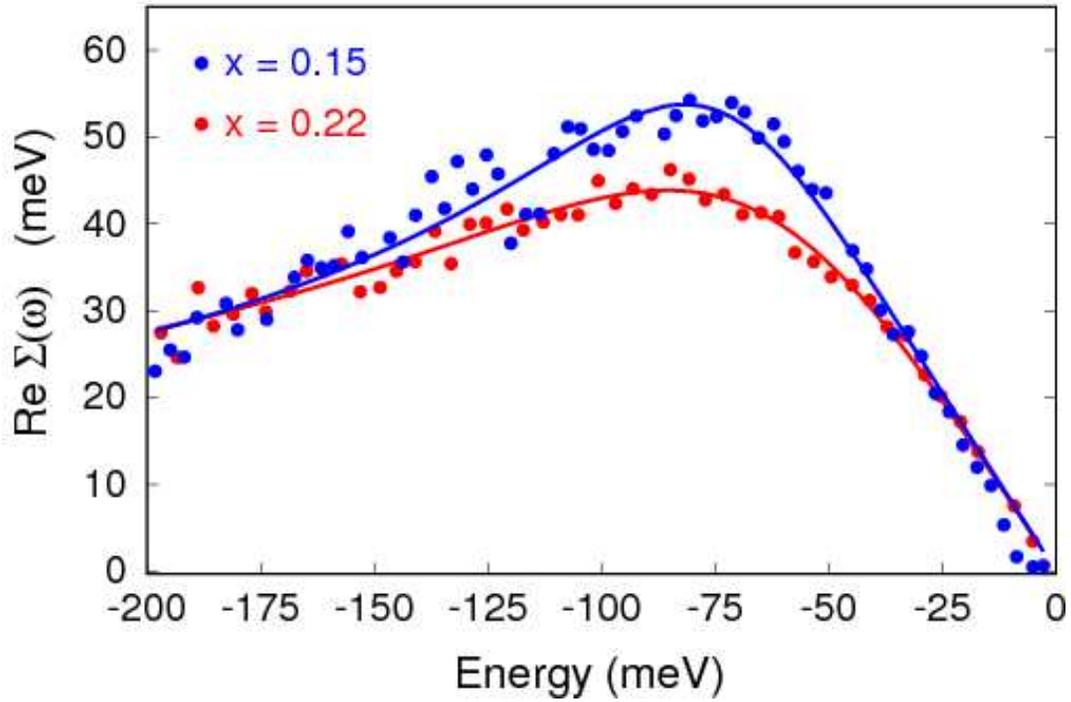

**Supplementary Figure 3. Extraction of the real part of the electron self-energy from the experimental data.** Data points extracted from Fig. 1(a) of ref 13 (optimally doped LSCO at 20 K, blue disks) and Fig. 1(d) of ref 14 (overdoped LSCO at 20 K, red disks). We determined the slope of the noninteracting dispersions by fitting the raw data to the model given by Eqs. (S2) and (S3). The solid lines indicate the model self energy curves obtained with this procedure.